\documentclass[twocolumn,showpacs,preprintnumbers,amsmath,amssymb]{revtex4}
\usepackage{dcolumn}% Align table columns on decimal point
\usepackage{bm}% bold math
\usepackage{graphicx}%
\usepackage{epsfig}
\begin{document}
\title{Nonequilibrium thermal entanglement  for simple qubit systems}
\author{Ilya Sinayskiy}
\email{sinayskiy@ukzn.ac.za} \affiliation{Quantum Research Group, School of Physics and
National Institute for Theoretical Physics, University of KwaZulu-Natal, Durban, 4001, South Africa}
\author{Nathan Pumulo}
\email{nate95@gmail.com}
\affiliation{Quantum Research Group, School of Physics and
National Institute for Theoretical Physics, University of KwaZulu-Natal, Durban, 4001, South Africa}
\author{Francesco Petruccione}
\email{petruccione@ukzn.ac.za}
\affiliation{Quantum Research Group, School of Physics and
National Institute for Theoretical Physics, University of KwaZulu-Natal, Durban, 4001, South Africa}


\begin{abstract}
The dynamics of simple qubit systems in a chain configuration coupled at both ends to
separate bosonic baths at different temperatures is studied. An
exact analytical solution of the master equation in the
Born-Markov approximation for the reduced density matrix of the
qubit system is constructed. The unique non-equilibrium stationary state for the long time behavior of the reduced density matrix in obtained. Dynamical and steady state properties of the concurrence between the first and the last spin are studied.
\end{abstract}

\maketitle

\section{Introduction}
In describing real physical systems one should always take into
account the influence of the surroundings. The study of open
systems is particularly important for understanding processes in
quantum physics \cite{toqs}. Usually, the unavoidable interaction with the environment destroys quantum correlations in the system. However, some times dissipation can be used not only for creation of entanglement in the system \cite{ilya,Huang,TE,braun}, but also for construction of non-trivial complex states of the system \cite{Z1,Z2}. An entanglement created by the interaction with thermal reservoir is called thermal entanglement. During the interaction with the thermal environment at temperature $T$ the system is evolving towards its steady state given by the Gibbs distribution $\rho\rightarrow e^{-H_S/k_BT}/\mathrm{Tr}[e^{-H_S/k_BT}]$  and sometimes this state is entangled. In this paper another kind of thermal entanglement will be discussed. Here, the system will be driven to the steady state by the interaction with two reservoirs at different temperatures \cite{ilya,Huang,Qui,SPP}

The paper is organized as follows. In Section II the
model of a qubit system in a chain configuration coupled at both ends to bosonic
baths at different temperatures is introduced.  The general strategy of derivation of the master equation for the reduced density matrix in the Born-Markov limit is presented. In Section III the two qubit case is considered. The master equation and its steady state solution are constructed. In Section IV the master equation for a three qubit case is obtained. The steady state solution in this case is constructed. Finally, in Section V results and conclusions are presented.

\section{Model}
We consider a qubit system in a chain configuration. The first and the last qubits of the chain are coupled to separate
bosonic baths at different temperatures. In this paper units are chosen in a such way that $\hbar=k_B=1$.The total Hamiltonian of
the system is given by
\begin{equation}
\hat{H}=\hat{H}_S+\hat{H}_{B1}+\hat{H}_{BN}+\hat{H}_{SB1}+\hat{H}_{SBN},
\end{equation}
where $\hat{H}_S$ is the Hamiltonian of the qubit subsystem,
\begin{equation}\label{HS}
\hat{H}_S=\sum_{i=1}^N\frac{\epsilon_i}{2}\hat{\sigma}^z_i+K\sum_{i=1}^{N-1}\left(\hat{\sigma}^+_i\hat{\sigma}^-_{i+1}+\hat{\sigma}^-_i\hat{\sigma}^+_{i+1}\right).
\end{equation}
The thermal reservoirs are described by an infinite set of harmonic oscillators, the Hamiltonians of the reservoirs coupled to the first qubit $(j=1)$ and to
the last qubit $(j=N)$ are given by
\begin{equation}
\hat{H}_{Bj}=\sum_n \omega_{n,j}\hat{b}^\dag_{n,j}\hat{b}_{n,j}.
\end{equation}
The interaction between the qubit subsystem and the bosonic baths
is described by the Jaynes Cummings type Hamiltonian in the rotating wave approximation
\begin{equation}
\hat{H}_{SBj}=\hat{\sigma}_{j}^{+}\sum_{n}g_{n}^{(j)}\hat{b}_{n,j}+\hat{\sigma}_{j}^{-}\sum_{n}g_{n}^{(j)*}\hat{b}_{n,j}^{\dag},
\label{Sinayskiy_eq4}
\end{equation}
of course, $\hat{\sigma}_{j}^{\pm},\hat{\sigma}_{j}^{z}$ are the
well-known Pauli matrices and $\hat{b}^\dag_{n,j},\hat{b}_{n,j}$
denote bosonic creation and annihilation operators.

The dynamics of the total system (qubits plus reservoirs) is described by the Liouville equation
\begin{equation}
 i\frac{d}{dt}\hat{\rho}(t)=[\hat{H},\hat{\rho}(t)].
\end{equation}
After performing the Born-Markov approximation~\cite{toqs} the equation
for the reduced density matrix (qubit subsystem only) in the interaction picture takes the following form:
\begin{eqnarray}
& &\frac{d}{dt}\hat{\rho}_S^{(I)}(t)=\\\nonumber & &-\int_0^\infty
ds\mathrm{tr}_{B}[\hat{H}_{SB}^{(I)}(t),[\hat{H}_{SB}^{(I)}(t-s),\hat{\rho}_S^{(I)}(t)\otimes\hat{\rho}_B(0)]],
\end{eqnarray}
where, the operator $\hat{\rho}_B(0)$ denotes the initial state of the reservoirs
\begin{equation}
\hat{\rho}_B(0)=\frac{e^{-\beta_1\hat{H}_{B1}}}{\mathrm{Tr}[e^{-\beta_1\hat{H}_{B1}}]}\otimes\frac{e^{-\beta_N\hat{H}_{BN}}}{\mathrm{Tr}[e^{-\beta_N\hat{H}_{BN}}]},
\end{equation}
and $\hat{H}_{SB}^{(I)}=\hat{H}_{SB1}^{(I)}+\hat{H}_{SBN}^{(I)}$.
After performing the rotating wave approximation over the rapidly
oscillating term in the master equation one gets:

\begin{eqnarray}\label{mseq}
\frac{d}{dt}\hat{\rho}_S(t)&=&-i[\hat{H}_S,\hat{\rho}_S(t)]+\\\nonumber
& &\sum_\omega\sum_{\alpha,\beta}\sum_{i,j}{\gamma_{\alpha,\beta}^{i,j}(\omega){\Big(}\hat{V}_\beta^i(\omega)\hat{\rho}_S(t)\hat{V}_{\alpha}^{j\dag}(\omega)}-\\\nonumber
& &\frac{1}{2}\left[\hat{V}_\alpha^{j\dag}(\omega)\hat{V}_\beta^i(\omega),\hat{\rho}_S(t)\right]_+\Big).
\end{eqnarray}

To obtain the master equation (\ref{mseq}) it assumed that the
system-environment interaction has the form 
\begin{equation}
\hat{H}_{SB_j}=\sum_\alpha\hat{V}_\alpha^{j\dag}\otimes\hat{f}_\alpha^j+\mathrm{h.c.},
\end{equation}
 the operators $\hat{V}_\alpha^j$ and $\hat{f}_\alpha^j$ acts on
the qubit system and the reservoir degrees of freedom, respectively. In the master equation
(\ref{mseq}) a \emph{Lamb-type} renormalization Hamiltonian was
neglected and decay rates $\gamma_{\alpha,\beta}^{i,j}(\omega)$ are
given by the Fourier image of the bath correlation functions:
\begin{equation}
\gamma_{\alpha,\beta}^{i,j}(\omega)=\int_{-\infty}^{+\infty}ds
e^{i\omega
s}\langle\hat{f}_{\alpha}^{i\dag}(s)\hat{f}_{\beta}^j(0)\rangle.
\end{equation}
On should note that, in Eq. (8) $\sum_{\omega}$ is a sum over all Bohr frequencies of the system. In Eqs. (8) and (10) indexes $\alpha$ and $\beta$ refer to decomposition of the interaction Hamiltonian into eigenoperators of the system Hamiltonian $\hat{H}_S$ and indexes $i$ and $j$ label the baths. 
The transition operators $\hat{V}_\alpha^j(\omega)$ originate from
the decomposition of the operator $\hat{V}_\alpha^j$ in the basis of
the eigenoperators of the system Hamiltonian $\hat{H}_S$. If one
denotes the eigenvalues of the Hamiltonian $\hat{H}_S$ by
$\varepsilon$ and corresponding projection operator as
$\hat{\Pi}(\varepsilon)$ then:
\begin{equation}
\hat{V}_\alpha^j(\omega)=\sum_{\varepsilon'-\varepsilon=\omega}\hat{\Pi}(\varepsilon)\hat{V}_\alpha^j\hat{\Pi}(\varepsilon').
\end{equation}
In other words each operator $\hat{V}_\alpha^j(\omega)$ satisfies the following equation:
\begin{equation}
[\hat{H}_S,\hat{V}_\alpha^j(\omega)]=-\omega\hat{V}_\alpha^j(\omega),
\end{equation}
where $\omega$ corresponding frequency of transition.

For the sake of simplicity and exact solvability in the paper we will considered only two cases: two and three qubit systems and symmetric Hamiltonian of the qubit system (\ref{HS}), i.e., for any $i$ all $\epsilon_i=\epsilon$.
\smallskip
\section{Two qubit system}
In the two qubit case the Hamiltonian of the qubit subsystem (\ref{HS}) takes a simple form 
\begin{equation}
\hat{H}_S=\frac{\epsilon}{2}\left(\hat{\sigma}^z_1+\hat{\sigma}^z_2\right)+K\left(\hat{\sigma}^+_1\hat{\sigma}^-_{2}+\hat{\sigma}^-_1\hat{\sigma}^+_{2}\right)
\end{equation}
and can easily be diagonalized with eigenevectors and corresponding eigenvalues given below
\begin{equation}
|m_1\rangle=|00\rangle,m_1=-\frac{\epsilon}{2},
\end{equation}
\begin{equation}
|m_2\rangle=|11\rangle,m_2=\frac{\epsilon}{2},
\end{equation}
\begin{equation}
|m_3\rangle=\frac{1}{\sqrt{2}}\left(|10\rangle+|01\rangle\right),m_3=K,
\end{equation}
\begin{equation}
|m_4\rangle=\frac{1}{\sqrt{2}}\left(-|10\rangle+|01\rangle\right),m_4=-K.
\end{equation}
In this case the master equation (\ref{mseq}) takes the following form
\begin{equation}\label{me2}
\frac{d\hat{\rho}}{dt}=-i[\hat{H}_{S},\hat{\rho}]+{\cal
L}_{1}(\hat{\rho})+{\cal L}_{2}(\hat{\rho}),\end{equation}
 where the superoperators ${\cal L}_i$ describe dissipation to the corresponding reservoir:
 \begin{eqnarray}
 {\cal L}_i(\hat{\rho})&=&\sum_{j=1}^2\gamma^{(i)}(-\omega_j^i)\Big(\hat{V}_j^i(\omega)\hat{\rho}\hat{V}_j^{i\dag}(\omega)\\\nonumber& &-\frac{1}{2}[\hat{V}_j^{i\dag}(\omega)\hat{V}_j^{i}(\omega),\hat{\rho}]_+\Big)\\\nonumber& &+\gamma^{(i)}(\omega_j^i)\Big(\hat{V}_j^{i\dag}(\omega)\hat{\rho}\hat{V}_j^{i}(\omega))\\\nonumber& &-\frac{1}{2}[\hat{V}_j^{i}(\omega)\hat{V}_j^{i\dag}(\omega),\hat{\rho}]_+\Big).
 \end{eqnarray}
The transition operators $\hat{V}_j^{i}(\omega)$ are
\begin{equation}
\hat{V}_1^1(\omega)=\frac{1}{\sqrt{2}}\left(|m_1\rangle\langle m_3|+|m_4\rangle\langle m_2|\right),
\end{equation}
\begin{equation}
\hat{V}_2^1(\omega)=\frac{1}{\sqrt{2}}(|m_3\rangle\langle m_2|-|m_1\rangle\langle m_4|),
\end{equation}
\begin{equation}
\hat{V}_1^2(\omega)=\frac{1}{\sqrt{2}}(|m_1\rangle\langle m_3|-|m_4\rangle\langle m_2|),
\end{equation}
\begin{equation}
\hat{V}_2^2(\omega)=\frac{1}{\sqrt{2}}(|m_3\rangle\langle m_2|+|m_1\rangle\langle m_4|),
\end{equation}
and the corresponding frequencies of trasition are
\begin{equation}
\omega_1^1=\omega_1^2=\omega_1=\epsilon-K,
\end{equation}
\begin{equation}
\omega_2^1=\omega_2^2=\omega_2=\epsilon+K.
\end{equation}
In this paper reservoirs are assumed to be represented by the infinite set of harmonic oscillators so that the decay rates read
\begin{equation}
\gamma^{(ij)}(\omega)\equiv\gamma^{(i)}(\omega)\delta_{ij}=\gamma_i n_i(\omega)\delta_{ij}=\frac{\gamma_i}{e^{\beta_i\omega}-1}\delta_{ij},
\end{equation}
\begin{equation}
\gamma^{(i)}(-\omega)=\gamma_i n_i(\omega)e^{\beta_i\omega},
\end{equation}
where $\gamma_i$ is the relaxation rate given by the spectral density of the reservoir.

The master equation (\ref{me2}) is solved exactly and details of the solution (including non-symmetric case $\epsilon_1\neq\epsilon_2$) can be found in \cite{ilya}. In this paper we will need only the non-equilibrium stationary state of the two qubit system for the comparison with the three qubit case. After some straightforward algebra the steady state of the two qubit system in the standard basis ${|00\rangle,|01\rangle,|10\rangle,|11\rangle}$ has the form:
\begin{widetext}
\[
\rho_{\infty}=\frac{1}{X_{1}X_{2}}\times
\]

\begin{equation}\label{dn2}
\left(\begin{array}{cccc}
X_{1}^{-}X_{2}^{-} &   0    &    0   &   0   \\
  0    & \frac{1}{2}\left(X_{1}^{-}X_{2}^{+}+X_{1}^{+}X_{2}^{-}\right) & \frac{1}{2}\left(X_{1}^{-}X_{2}^{+}-X_{1}^{+}X_{2}^{-}\right) &   0   \\
  0    & \frac{1}{2}\left(X_{1}^{-}X_{2}^{+}-X_{1}^{+}X_{2}^{-}\right) & \frac{1}{2}\left(X_{1}^{-}X_{2}^{+}+X_{1}^{+}X_{2}^{-}\right) &   0   \\
  0    &   0    &    0   & X_{1}^{+}X_{2}^{+}\end{array}\right),
\end{equation}
\end{widetext}
where $X_{i}^{\mp} = \gamma^{(1)}(\pm\omega_{i})+\gamma^{(2)}(\pm\omega_{i})$ and $X_{i}=X_{i}^{+}+X_{i}^{-}$.
In order to quantify the entanglement between the qubits we
consider the concurrence~\cite{woot}. In the steady-state
$(t\rightarrow\infty)$ it is given by
\begin{widetext}
\begin{equation}
C_{\infty}=\frac{2}{X_1X_2}\textrm{Max}\left(0,\frac{1}{2}|X_1^+X_2^--X_1^-X_2^+|-\sqrt{X_1^-X_1^+X_2^-X_2^+}\right)
.\end{equation}
\end{widetext}

\smallskip

\section{Three qubit system}
The three qubit case is also exactly solvable. Following the same strategy one can find eigenvalues and eigenvectors for the Hamiltonian $\hat{H}_S$ (\ref{HS}) in the case of three qubits:
\begin{equation}
|m_1\rangle=|000\rangle,
\end{equation}
\begin{equation}
|m_2\rangle=\frac{|001\rangle-|100\rangle}{\sqrt{2}},
\end{equation}
\begin{equation}
|m_3\rangle=\frac{|011\rangle-|110\rangle}{\sqrt{2}},
\end{equation}
\begin{equation}
|m_4\rangle=|111\rangle,
\end{equation}
\begin{equation}
|m_5\rangle=\frac{|100\rangle-\sqrt{2}|010\rangle+|001\rangle}{2},
\end{equation}
\begin{equation}
|m_6\rangle=\frac{|110\rangle-\sqrt{2}|101\rangle+|011\rangle}{2},
\end{equation}
\begin{equation}
|m_7\rangle=\frac{|100\rangle+\sqrt{2}|010\rangle+|001\rangle}{2},
\end{equation}
\begin{equation}
|m_8\rangle=\frac{|110\rangle+\sqrt{2}|101\rangle+|011\rangle}{2},
\end{equation}
and corresponding eigenvalues $m_i$:
\begin{equation}
m_1=-m_4=-\frac{3\epsilon}{2},
\end{equation}
\begin{equation}
m_2=-m_3=-\frac{\epsilon}{2},
\end{equation}
\begin{equation}
m_5=-m_8=-\frac{\epsilon}{2}-\sqrt{2}K,
\end{equation}
\begin{equation}
m_6=-m_7=\frac{\epsilon}{2}-\sqrt{2}K.
\end{equation}

In this case the master equation (\ref{mseq}) for the reduced density matrix reads
\begin{equation}\label{me3}
\frac{d\hat{\rho}}{dt}=-i[\hat{H}_{S},\hat{\rho}]+{\cal
L}_{1}(\hat{\rho})+{\cal L}_{3}(\hat{\rho})\end{equation}
 where the superoperators ${\cal L}_i$ are given by
  \begin{eqnarray}
 {\cal L}_i(\hat{\rho})&=&\sum_{j=1}^3\gamma^{(i)}(-\omega_j^i)\Big(\hat{V}_j^i(\omega)\hat{\rho}\hat{V}_j^{i\dag}(\omega)\\\nonumber& &-\frac{1}{2}[\hat{V}_j^{i\dag}(\omega)\hat{V}_j^{i}(\omega),\hat{\rho}]_+\Big)\\\nonumber& &+\gamma^{(i)}(\omega_j^i)\Big(\hat{V}_j^{i\dag}(\omega)\hat{\rho}\hat{V}_j^{i}(\omega))\\\nonumber& &-\frac{1}{2}[\hat{V}_j^{i}(\omega)\hat{V}_j^{i\dag}(\omega),\hat{\rho}]_+\Big).
 \end{eqnarray}
There are three transition operators $\hat{V}_j^{i}(\omega)$ for each reservoir, namely,
\begin{widetext}
\begin{equation}
\hat{V}_1^1(\omega)=\frac{1}{\sqrt{2}}\left(-|m_1\rangle\langle m_2|+|m_3\rangle\langle m_4|-|m_5\rangle\langle m_6|+|m_7\rangle\langle m_8|\right),
\end{equation}
\begin{equation}
\hat{V}_2^1(\omega)=\frac{1}{2}\left(| m_1\rangle\langle m_5|-| m_2\rangle\langle m_6|-| m_7\rangle\langle m_3|+| m_8\rangle\langle m_4|\right),
\end{equation}
\begin{equation}
\hat{V}_3^1(\omega)=\frac{1}{2}\left(| m_1\rangle\langle m_7|+| m_2\rangle\langle m_8|+| m_5\rangle\langle m_3|+| m_6\rangle\langle m_4|\right),
\end{equation}
\begin{equation}
\hat{V}_1^3(\omega)=\frac{1}{\sqrt{2}}\left(| m_1\rangle\langle m_2|-| m_3\rangle\langle m_4|-| m_5\rangle\langle m_6|+| m_7\rangle\langle m_8|\right),
\end{equation}
\begin{equation}
\hat{V}_2^3(\omega)=\frac{1}{2}\left(| m_1\rangle\langle m_5|+| m_2\rangle\langle m_6|+| m_7\rangle\langle m_3|+| m_8\rangle\langle m_4|\right),
\end{equation}
\begin{equation}
\hat{V}_3^3(\omega)=\frac{1}{2}\left(| m_1\rangle\langle m_7|-| m_2\rangle\langle m_8|-| m_5\rangle\langle m_3|+| m_6\rangle\langle m_4|\right),
\end{equation}
\end{widetext}
and three corresponding frequencies of transition
\begin{equation}
\omega_1^1\equiv\omega_1^3\equiv\omega_1=\epsilon,
\end{equation}
\begin{equation}
\omega_2^1\equiv\omega_2^3\equiv\omega_2=\epsilon-\sqrt{2}K,
\end{equation}
\begin{equation}
\omega_3^1\equiv\omega_3^3\equiv\omega_3=\epsilon+\sqrt{2}K,
\end{equation}
please note that $\gamma^{(i)}(\pm\omega_j^i)$ has the same meaning like in the two-qubit case and is given by Eqs. (26) and (27).

The master equation (\ref{me3}) is solved exactly. The details of the solution can be found in \cite{SPP}. One can study the dynamics of the concurrence of the system for a certain initial state of the system. In Figure 1 the dynamics of the concurrence between first and third qubit for initial $|W_3\rangle$ -state is presented
\begin{equation}
|W_3\rangle=\frac{1}{\sqrt{3}}\left(|100\rangle+|010\rangle+|001\rangle\right).
\end{equation}
For the all three cases presented in Figure 1 one can identify two time scales of the dynamics, the first timescale is the time of dissipation of the entanglement created by the $XX$-interaction between qubits and the second timescale is the creation of the thermal entanglement. The phenomena of disappearing and reappearing of the entanglement is called sudden death and sudden birth of entanglement \cite{eberly}.

Using the exact solution of the master equation (\ref{me3}) one can find the long-time behavior of the density matrix. It is possible to show that in the basis of eigenvectors $\{|m_i\rangle\}$ of the qubit  Hamiltonian $\hat{H}_S$ the non-equlibrium steady-state of the reduced density matrix of the qubit system will be diagonal

\begin{equation}\label{dn3}
\rho^{ii}_\infty=\frac{1}{X_1X_2X_3}\left(\begin{array}{c}
X_1^+X_2^+X_3^+\\
X_1^-X_2^+X_3^+\\
X_1^+X_2^-X_3^-\\
X_1^-X_2^-X_3^-\\
X_1^+X_2^-X_3^+\\
X_1^-X_2^-X_3^+\\
X_1^+X_2^+X_3^-\\
X_1^-X_2^+X_3^-\end{array}\right),
\end{equation}
where $X_{i}^{\pm} = \gamma^{(1)}(\pm\omega_{i})+\gamma^{(3)}(\pm\omega_{i})$ and $X_{i}=X_{i}^{+}+X_{i}^{-}$.

Using this non-equilibrium steady state one can analyze the concurrence in the system. In Figure 2 and 3 a comparison of the steady state concurrence in the two qubit and three qubit case is presented. Figure 2 addresses the dependence of the steady state concurrence form the reservoir temperatures in the equilibrium case. Figure 3 presents the dependence of the steady state concurrence in the non-equilibrium case. From both figures one can see that there is an interval of temperatures for which the steady state concurrence for three qubit system is higher than in the two qubit case.

%-------------
\begin{figure}
\centering
\includegraphics[width=8.25 cm]{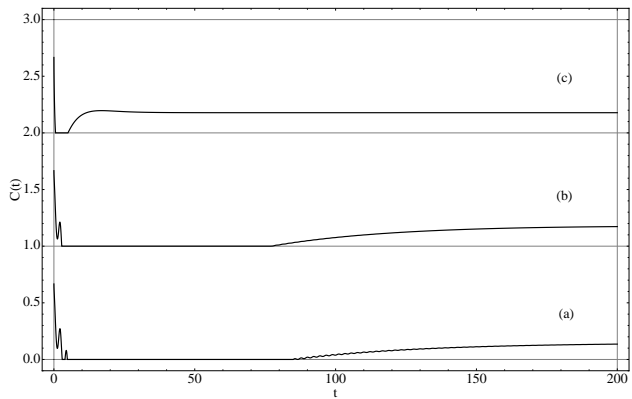}
%\vspace{0.2cm}
\caption{Dynamics of the concurrence between the first and the third qubit for different temperatures
of the reservoirs. The initial state of the three qubit system is the 
$|W_3\rangle$ state. Curve $(a)$ corresponds to
$\beta_1=\beta_3=10$, curve $(b)$ to $\beta_1=\beta_3=5$, curve $(c)$ to $\beta_1=5$, $\beta_3=3$; all
the other parameters are the same:
$\epsilon=3/2$,$\kappa=1$,$\gamma_1=\gamma_N=1/50$. For a convenience curves $(b)$ and $(c)$ are shifted up on the Y axes by 1 and 2 units, respectively.}
\end{figure}

\begin{figure}
\centering
\includegraphics[width=8.25 cm]{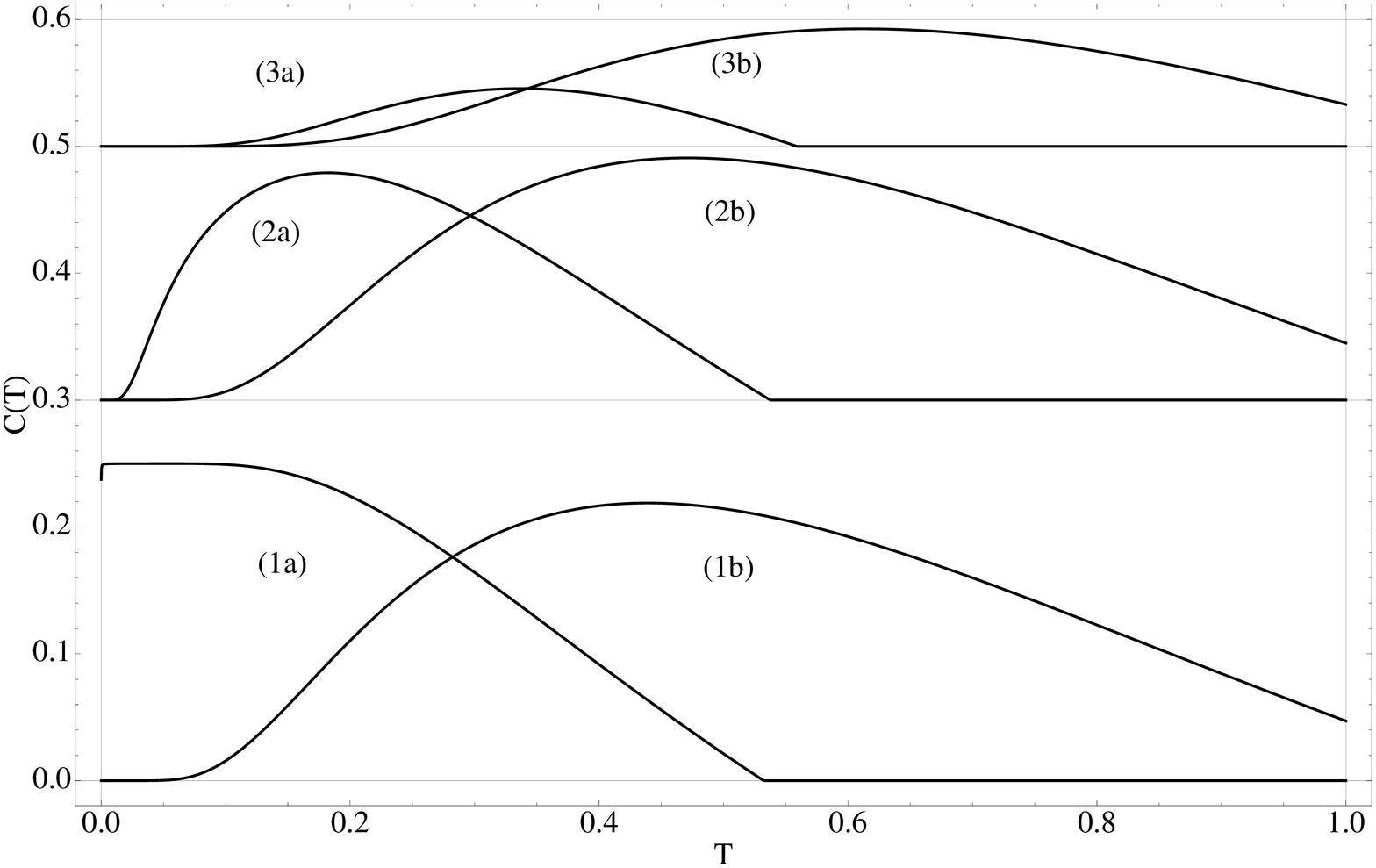}
%\vspace{0.2cm}
\caption{Comparison of the steady-state entanglement for two qubit
and three qubit systems in the thermal equilibrium. Curves $(1a)$, $(2a)$ and $(3a)$
correspond to the three qubit system and curves $(1b)$, $(2b)$ and $(3b)$ to the two qubit system. Curves $(1a)$ and $(1b)$ correspond to $\epsilon/K=\sqrt 2$, curves $(2a)$ and $(2b)$ correspond to $\epsilon/K=3/2$ and curves $(3a)$ and $(3b)$ correspond to $\epsilon/K=2$.  For a convenience curves $2$ and $3$ are shifted up on the Y axes by 0.3 and 0.5 units, respectively.}
\end{figure}

\begin{figure}
\centering
\includegraphics[width=8.25 cm]{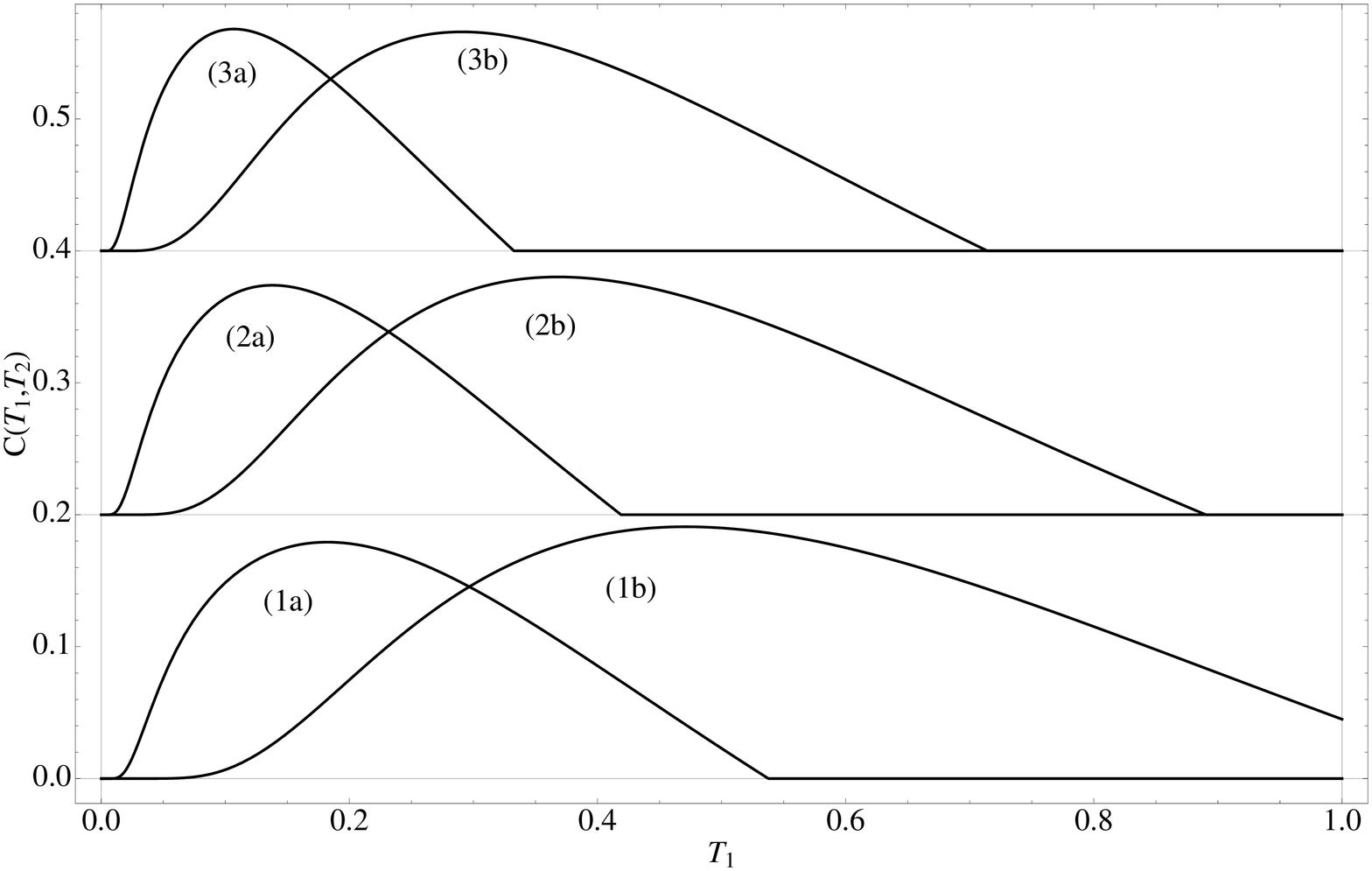}
%\vspace{0.2cm}
\caption{Comparison of the steady-state entanglement for two qubit
and three qubit systems in the non-equilibrium case. Curves $(1a)$, $(2a)$ and $(3a)$
correspond to the three qubit system and curves $(1b)$, $(2b)$ and $(3b)$ to the two qubit system. Curves $(1a)$ and $(1b)$ correspond to equilibrium case $T_2=T_1$, curves $(2a)$ and $(2b)$ correspond to $T_2=3T_1/2$, curves $(3a)$ and $(3b)$ correspond to $T_2=2T_1$. The other parameters are $\epsilon=3/2$ and $K=1$.  For a convenience curves $2$ and $3$ are shifted up on the Y axes by 0.2 and 0.4 units respectively.}
\end{figure}
%-------------
\section{Results and Conclusion}
It is important to note that the non-equilibrium steady state for two qubit (\ref{dn2}) and three qubit (\ref{dn3}) systems presented here are entangled states. In the equilibrium case $(T_1=T_2)$ the steady states takes the form of the Gibbs state
\begin{equation}
\hat{\rho}_\infty=\frac{e^{-\beta\hat{H}_S}}{\mathrm{Tr}[e^{-\beta\hat{H}_S}]}.
\end{equation} 
In Figure 1 one can see the system approaching equilibrium. For the reservoirs of low temperature (curves $a$ and $b$) one can see the competition between exchange $XX$-interaction and irreversible dissipative dynamics in the system evolution. One can also see entanglement sudden birth and sudden death. In the case of the high temperature (curve $c$) one can see only exponential decay of entanglement and after some time sudden birth of thermal entanglement.

In Figure 2 and 3 a comparison of the steady state concurrence between two and three qubit systems is performed. In the case of the three qubit system the concurrence between first and third qubit is studied. In Figure 2 one can see that increasing the $\epsilon/K$ ratio the amount of concurrence decreased for the both systems. But for all considered $\epsilon/K$ ratios there are intervals of the temperatures for which the amount of the steady state concurrence in the three qubit case is higher that in two qubit one.

In Figure 3 we show a comparison between steady state concurrence for two and three qubit system in the non-equilibrium case. One can see that in all cases there are intervals of the temperatures of the reservoirs for which the steady state concurrence in the three qubit case is higher than in a two qubit one. But as in the symmetric two
qubit case the steady state entanglement reaches
its maximal value in the equilibrium case.

In conclusion, we have presented an analytical expression for a two and three qubit
system in a chain configuration coupled to bosonic baths at different temperatures. The dynamics of the system is studied and it is shown that the system convergences to a
non-equilibrium steady state. The dynamics of entanglement is analyzed and a
comparison of the steady state concurrence of two and
three qubit systems is performed. It is found that there is a range of parameters in which the
three qubit system contains more quantum correlations in the steady
state than two qubit one.

This work is based upon research supported by the South African
Research Chair Initiative of the Department of Science and
Technology and National Research Foundation.


\begin{thebibliography}{0}
\bibitem{toqs} H.-P.Breuer and F.Petruccione, The Theory
of Open Quantum Systems (Oxford University Press, Oxford, 2002).
\bibitem{ilya}I. Sinaysky, F, Petruccione, and D. Burgarth,
Phys. Rev. A {\bf 78} 062301 (2008).
\bibitem{Huang} X. L. Huang, J. L. Guo, X. X. Yi, Phys. Rev. \textbf{A80}, 054301,
(2009).
\bibitem{TE} M. C. Arnesen et al., Phys. Rev. Lett. \textbf{87}, 017901 (2001); X.
G. Wang, Phys. Rev. \textbf{A64}, 012313 (2001); X. G.Wang et al.,
J. Phys. A \textbf{34}, 11307 (2001); X. G. Wang, Phys. Rev. A
\textbf{66}, 034302 (2002); X. G. Wang, ibid. \textbf{66}, 044305
(2002); G. L. Kamta and A. F. Starace, Phys. Rev. Lett.
\textbf{88}, 107901 (2002); L. Zhou et al., Phys. Rev.
\textbf{A68}, 024301 (2003); Y. Sun et al., ibid. \textbf{68},
044301 (2003); M. Cao and S. Zhu, ibid. \textbf{71}, 034311
(2005).
\bibitem{braun} M.B. Plenio and S. F. Huelga, Phys. Rev. Lett., \textbf{88}, 197901 (2002); D. Braun, {\it ibid}. \textbf{89}, 277901
(2002).
\bibitem{Z1} S. Diehl, A. Micheli, A. Kantian, B. Kraus, H. P. B\"uchler, P. Zoller, Nature Physics \textbf{4}, 878 (2008)
\bibitem{Z2} B. Kraus, H. P. B\"uchler, S. Diehl, A. Kantian, A. Micheli, P. Zoller,  Phys. Rev. A   {\bf 78} 042307 (2008)
\bibitem{Qui} L. Quiroga, F.J. Rodriguez, M.E. Ramirez, R. Paris, Phys. Rev. \textbf{A75}, 032308
(2007).

\bibitem{SPP}I. Sinayskiy, N. Pumulo, F, Petruccione, \textit{(to be published)} (2011).
\bibitem{woot} W.K. Wootters, Phys. Rev. Lett., (1998), \textbf{80},
2245-2248.
\bibitem{eberly} T. Yu and J. H. Eberly, Phys. Rev. Lett. \textbf{93}, 140404
(2004)

\end{thebibliography}
\end{document}